\documentclass[twocolumn,showpacs,preprintnumbers,amsmath,amssymb,superscriptaddress]{revtex4}
\usepackage{graphicx} % Include figure files
\graphicspath{{fig/}}
\usepackage{dcolumn}  % Align table columns on decimal point
\usepackage{bm}       % bold math
\usepackage{natbib}
\usepackage{epsfig}
\usepackage{amssymb}
\usepackage{multirow}
\usepackage{amssymb}
\usepackage{graphicx}
\usepackage{color}

\begin{document}
\title{Response of liquid xenon to Compton electrons down to 1.5\,keV}
\author{L.~Baudis}\affiliation{Physics Institute, University of Zurich, Winterthurerstr. 190, 8057 Zurich, Switzerland}
\author{H.~Dujmovic}\affiliation{Physics Institute, University of Zurich, Winterthurerstr. 190, 8057 Zurich, Switzerland}
\author{C.~Geis}\thanks{Present address: Institut f\"ur Physik, Johannes Gutenberg Universit\"at Mainz, 55099 Mainz, Germany}\affiliation{Physics Institute, University of Zurich, Winterthurerstr. 190, 8057 Zurich, Switzerland}\affiliation{Technische Hochschule Mittelhessen, Wilhelm-Leuschner-Strasse 13, 61169 Friedberg, Germany}
\author{A.~James}\affiliation{Physics Institute, University of Zurich, Winterthurerstr. 190, 8057 Zurich, Switzerland}
\author{A.~Kish}\affiliation{Physics Institute, University of Zurich, Winterthurerstr. 190, 8057 Zurich, Switzerland}
\author{A.~Manalaysay}\email[Corresponding author: ]{aaronm@physik.uzh.ch}\affiliation{Physics Institute, University of Zurich, Winterthurerstr. 190, 8057 Zurich, Switzerland}
\author{T.~Marrod\'an Undagoitia}\thanks{Present address: Max-Plank-Institut f\"ur Kernphysik, Saupfercheckweg 1, 69117 Heidelberg, Germany}\affiliation{Physics Institute, University of Zurich, Winterthurerstr. 190, 8057 Zurich, Switzerland}
\author{M.~Schumann}\thanks{Present address: Albert Einstein Center for Fundamental Physics, University of Bern, Sidlerstr. 5, 3012 Bern, Switzerland}\affiliation{Physics Institute, University of Zurich, Winterthurerstr. 190, 8057 Zurich, Switzerland}

\begin{abstract}
The response of liquid xenon to low-energy electronic recoils is relevant in the search for dark-matter candidates which interact predominantly with atomic electrons in the medium, such as axions or axion-like particles, as opposed to weakly interacting massive particles which are predicted to scatter with atomic nuclei.  Recently, liquid-xenon scintillation light has been observed from electronic recoils down to 2.1\,keV, but without applied electric fields that are used in most xenon dark matter searches.  Applied electric fields can reduce the scintillation yield by hindering the electron-ion recombination process that produces most of the scintillation photons.  We present new results of liquid xenon's scintillation emission in response to electronic recoils as low as 1.5\,keV, with and without an applied electric field.  At zero field, a reduced scintillation output per unit deposited energy is observed below 10\,keV, dropping to nearly 40\% of its value at higher energies.  With an applied electric field of 450\,V/cm, we observe a reduction of the scintillation output to about 75\% relative to the value at zero field.  We see no significant energy dependence of this value between 1.5\,keV and 7.8\,keV.  With these results, we estimate the electronic-recoil energy thresholds of ZEPLIN-III, XENON10, XENON100, and XMASS to be 2.8\,keV, 2.5\,keV, 2.3\,keV, and 1.1\,keV, respectively, validating their excellent sensitivity to low-energy electronic recoils.
\end{abstract}

\maketitle

\section{Introduction}
\label{sec:intro}

Liquid xenon (LXe) provides an ideal detection medium for Weakly Interacting Massive Particles (WIMPs), which are promising and testable candidates for cold dark matter in the Milky Way \cite{Bertone:2004pz}.  As a noble liquid, xenon is both a very good scintillator and ionizer in response to the passage of radiation \cite{Aprile:2009dv}. The possibility to detect charge and light signals after a WIMP scatters on a xenon nucleus, along with the relative ease to scale-up to large masses and its self-shielding properties (e.g.~high stopping power for penetrating radiation) has made LXe the WIMP target of choice for many attempts to directly observe a WIMP-induced signal in the laboratory \cite{Angle:2007uj,Lebedenko:2008gb,Aprile:2011hi,Akerib:2011ix,Abe:2012az}.  

The expected signature of WIMP interactions in a LXe detector are low-energy depositions by the recoiling xenon nuclei up to several tens of keV.  A fraction of the energy of the recoiling nucleus is transferred to electronic excitations of the medium and is observable as ionization and scintillation.  The energy dependence of the ionization and scintillation signals in response to nuclear recoils has been investigated by multiple groups below 10\,keV \cite{Chepel:06,Aprile:2008rc,Manzur:2009hp,Sorensen:2011bd}, and most recently down to 3\,keV \cite{Plante:2011hw}.  In contrast, until recently \cite{Manalaysay:2009yq,Aprile:2012an}, studies of LXe's response to low-energy \emph{electronic} recoils have not been pursued with the same voracity that has accompanied the measurements of nuclear recoils.  This is not surprising, as electronic recoils present only background in searches for WIMPs.  However, dark matter searches that focus on WIMPs can also have sensitivity to non-WIMP dark-matter candidates that may interact with electrons.  For example, axion-like pseudoscalars, which couple to two photons, could ionize an atom via the axioelectric effect, which is analogous to the photoelectric process \cite{Avignone:1987,Aalseth:2010vx}.  This process has been considered as a possible dark matter interpretation of the annual modulation signal observed by the DAMA/LIBRA experiment in the region around 2-5\,keV \cite{Bernabei:2010mq,Bernabei:2005ca}.  Part of the parameter space consistent with this interpretation has been excluded by the CoGeNT and CDMS-II experiments \cite{Aalseth:2010vx,Ahmed:2009ht}.  LXe could, in principle, have sensitivity to much (if not all) of the remaining parameter space, but a measurement of LXe's response to low-energy electronic recoils under the conditions present in most LXe dark matter searches (i.e.~applied electric fields) has yet to be achieved.  

\begin{figure*}[ht!]
	\begin{center}
        \includegraphics[width=0.95\textwidth]{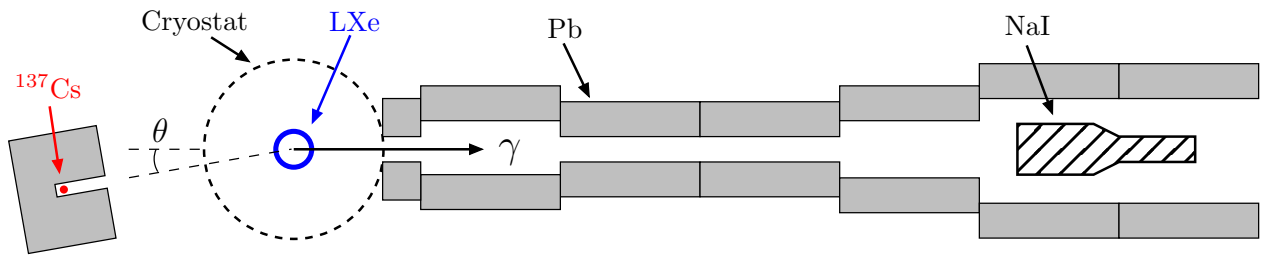} 
        \caption{Schematic top-view of the experimental setup.  The 662\,keV $\gamma$ rays are collimated twice: first as they leave the $^{137}$Cs source, and second after they scatter in the LXe volume.  The Pb channel from LXe to NaI is also covered on top and bottom (not shown).  The scattering angle, $\theta$, is varied from 4.25$^{\circ}$ to 34.5$^{\circ}$.}
        \label{fig:exp_setup}
    \end{center}
\end{figure*}

We report here a study of LXe's scintillation yield to electronic recoils as low as 1.5\,keV resulting from low-angle Compton scatters of 662\,keV $\gamma$ rays.  For this purpose, a small LXe cylindrical cell is irradiated with these $\gamma$ rays, a fraction of which are absorbed by a NaI scintillating crystal placed at a set of chosen angles relative to the original direction of the $\gamma$ rays, thus kinematically determining the energy of the recoiling electrons.  Recently, another group has used a similar technique to measure the LXe scintillation yield down to 2.1\,keV \cite{Aprile:2012an}.  We additionally measure the scintillation quenching induced by the application of a uniform electric field to the LXe, which is essential to infer the true energy thresholds of existing LXe dark matter searches, many of which apply fields of similar strength \cite{Angle:2007uj,Lebedenko:2008gb,Aprile:2011hi,Akerib:2011ix}.  This study represents the first observation of scintillation signals (both with and without applied fields) in this energy range.  The paper is organized as follows: in Section \ref{sec:methods} we describe the experimental methods used for the Compton scattering experiment, namely the LXe cell and the NaI detector.  Section \ref{sec:mc} describes the Monte Carlo methods used in simulations of the setup, and in Section \ref{sec:results} we present the data analysis, including comparison with detailed Monte Carlo simulations, and give the results of our measurements. In Section \ref{sec:discussion} we present a summary of our main findings, as well as a discussion and implications of the results for dark matter searches.

\section{Experimental Methods}
\label{sec:methods}

The Compton-scatter setup consists of a collimated $^{137}$Cs source, a small LXe scintillation cell, and a NaI scintillating crystal, shown schematically in Figure \ref{fig:exp_setup}.  The 17.3\,MBq $^{137}$Cs source emits 662\,keV $\gamma$ rays and is encased in a lead block with a small cylindrical opening, 0.6\,cm in diameter and 5\,cm long, that acts as a collimator.  Monte Carlo (MC) simulations of this source show that the resulting beam from the collimator has a 1\,$\sigma$ angular spread of 1.6$^{\circ}$.  The LXe cell, which is described in detail in \cite{Manalaysay:2009yq,AManalaysayPhD}, consists of a cylinder of LXe, 4.5\,cm tall and 3.5\,cm diameter, viewed on top and bottom by two 2''-diameter Hamamatsu R6041 photomultiplier tubes (PMTs), and surrounded by a polytetrafluoroethylene (PTFE) shell.  The PTFE acts as an efficient light reflector \cite{Yamashita2004692} which permits photons hitting the detector walls to still be detected in the PMTs.  Three flat grid electrodes, located at 0.5\,cm (cathode), 3.5\,cm (gate), and 4\,cm (anode) above the bottom photocathode, intersect the LXe cylinder and are used to apply static electric fields across the volume, parallel to the cylinder's axis.  In order to maximize the efficiency for detecting scintillation photons, LXe is filled fully from the bottom PMT to the top PMT, producing a single-phase detector.  This contrasts with most LXe dark matter detectors which use a dual-phase design in order to also detect very small ionization signals \cite{bolozdynyaDP:1999}; the scintillation signal in the present detector is reduced by $\sim$40\% when the liquid-gas interface is lowered below the top PMT.  The PMT photocathodes are held at ground potential, with positive high voltage applied to their anodes.  Throughout the run, the LXe is continuously recirculated and purified through a SAES Monotorr hot getter, in order to remove any impurities that may enter the liquid.  The NaI detector is a Saint-Gobain model 3M3/3, which is a fully integrated crystal and PMT.  The NaI crystal itself is a cylinder, 7.6\,cm in diameter and in 7.6\,cm height.

The opening of the source collimator is placed initially 70\,cm from the center of the LXe cell.  For a subset of the scattering angles (4.25$^{\circ}$, 5.25$^{\circ}$, and 8.5$^{\circ}$) this distance is reduced to 28\,cm (the minimum allowed given the detector components) in order to minimize the beam's spot size within the LXe volume.  A distance of $\sim$1\,m is chosen for the NaI position as a compromise between event rate, which decreases with larger separations, and angular systematics (see Section \ref{sec:mc}), which improves with increased separation.  The three components are aligned using a goniometer with 0.25$^{\circ}$ tick marks; this tick-mark width is taken to be the 1\,$\sigma$ accuracy ($\pm$0.125$^{\circ}$) of the geometrical alignment and is included as a systematic uncertainty in the analysis (see Section \ref{sec:results}).  The precision with which a scattering angle can be reproduced is better than the spacing between adjacent tick marks, and therefore associating this width as a 1$\sigma$ uncertainty is conservative.  Unless otherwise specified, reported scattering angles refer to the angle formed by the collimated beam with the centers of the detector components. After scattering in the LXe cell, the $\gamma$ rays are further collimated on their way to the NaI detector by means of a lead channel with a 3\,cm circular aperture at its entrance (LXe side), which then widens to encompass the NaI crystal and PMT (see Figure \ref{fig:exp_setup}).  Data are collected at central scattering angles of 4.25$^{\circ}$, 5.25$^{\circ}$, 6.25$^{\circ}$, 8.5$^{\circ}$, 16.25$^{\circ}$, and 34.5$^{\circ}$.  These correspond to expected electron energies of 2.35\,keV, 3.57\,keV, 5.05\,keV, 9.28\,keV, 32.5\,keV, and 123\,keV, respectively, when applying the well known Compton scatter formula,
\begin{equation}
\label{eq:compt}
E_{\mathrm{er}} = E_{\gamma}^2~ \frac{1-\mathrm{cos}\,\theta}{m_\mathrm{e}c^2 + E_{\gamma} (1-\mathrm{cos}\,\theta)} ~, 
\end{equation}
where $E_{\mathrm{er}}$ is the energy of the recoiling electron, $E_{\gamma}$ is the initial energy of the incident $\gamma$ ray, $m_{\mathrm{e}}$ is the mass of the electron, and $\theta$ is the scattering angle.  However, as will be shown in Section \ref{sec:mc}, the finite size of the detector components lead to peak recoil energies that differ from these expectations.

\begin{figure}[hbp!]
	\begin{center}
        \includegraphics[width=0.48\textwidth]{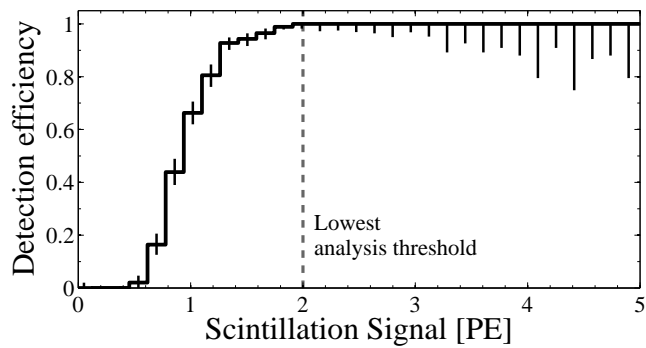} 
        \caption{The measured hardware detection efficiency of the experimental setup, as a function of the measured scintillation size in the LXe.  The lowest analysis threshold considered is 2\,PE, where the detection efficiency is already unity.}
        \label{fig:effic}
    \end{center}
\end{figure}

All three PMT signals---two from the LXe and one from the NaI---are read out directly, without amplification.  The signals are each split with a CAEN N625 linear Fan-out.  One triplet of signals from the Fan-out are passed to the trigger and timing system, while the other triplet is fed directly into an Acqiris DC436 100\,MS/s waveform digitizer.  The signals going to the trigger system are first passed to a CAEN N840 leading edge discriminator, where they are converted to logical signals.  Individual trigger thresholds for the two LXe channels are set at $\sim$1 photoelectron (PE).  The global trigger condition requires time coincidence between the NaI signal and \emph{either} of the two LXe signals.  In this way, the system utilizes a two-fold PMT coincidence requirement globally, but allows the LXe itself to trigger with only a single channel, thus optimizing the LXe scintillation detection efficiency.  The trigger detection efficiency is directly measured with a $^{22}$Na source, similar to the technique used in \cite{Plante:2011hw}.  The resulting detection efficiency curve is shown in Figure \ref{fig:effic}, reaching unity by $\sim$2\,PE, which is the smallest signal considered for analysis (Section \ref{sec:results}).

The logical signals from the LXe and NaI are additionally connected to the ``start'' and ``stop'' inputs of an Ortec 566 Time to Amplitude Converter (TAC).  Its output is digitized, along with the three PMT signals, to measure the Time of Flight (ToF) of the $\gamma$ ray and further constrain the time coincidence of the LXe and NaI signals offline.  The TAC module has an intrinsic timing resolution of better than 5\,ps, however, the true precision of the ToF measurement is limited by the rise time of the NaI scintillation pulse, which is of $\mathcal{O}$(ns).  The ToF measurement method is calibrated with the use of a $^{22}$Na source placed between the LXe and NaI.  This source undergoes $\beta^{+}$ decay, eventually emitting two 511\,keV photons (in opposite directions) when the positron loses energy and annihilates with an electron; a signal in both detectors physically indicates ToF$=0$ and can be shifted with variable delay generator to calibrate the full range of the TAC system.

\begin{figure}[htp!]
	\begin{center}
        \includegraphics[width=0.48\textwidth]{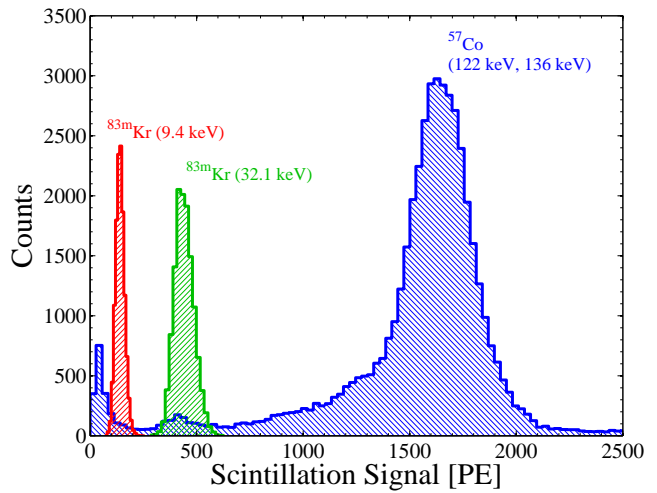} 
        \caption{(color online) Measured spectra in the LXe from the decay of $^{57}$Co (122\,keV and 136\,keV) and $^{83\mathrm{m}}$Kr (32.1\,keV and 9.4\,keV).  The feature in the $^{57}$Co spectrum at and below $\sim$500 PE is the Compton continuum from this source.}
        \label{fig:CoAndKr}
    \end{center}
\end{figure}
The PMT gains and light-yield stability are calibrated weekly, throughout the duration of the data taking.  The LXe PMT gains are measured by means of a low-intensity, pulsed, blue light emitting diode placed in the LXe but outside of the PTFE shell.  This placement is chosen in order to help diffuse the light reaching the PMTs, thereby illuminating all regions of the photocathodes.
%The intensity of the LED is reduced so that $\lesssim\!5$\% of LED pulses produce no associated pulse from the PMT; in this way, the collected PMT pulses contain a nearly pure sample of single photoelectrons (PE), having a two-PE contamination of $\lesssim\!2.5$\%.  
The procedure is similar to that used in experiments such as Borexino \cite{Dossi:1998zn} and XENON100 \cite{Aprile:2011dd}.  With the measured gains, all PMT signals from the LXe are converted to units of PE.  The light yield stability is monitored using two radioactive isotopes, $^{57}$Co (external, 122\,keV and 136\,keV $\gamma$ rays) and $^{83\mathrm{m}}$Kr (internal, 32.1\,keV and 9.4\,keV transitions, mostly conversion electrons).  The spectra from one set of calibrations with these sources are seen in Figure~\ref{fig:CoAndKr}.  The two $\gamma$ rays emitted by the $^{57}$Co are not separately distinguishable, and instead produce a broadened peak whose mean value inside the LXe, determined by MC simulations, is 126.1\,keV.  The LY obtained from this source is $(13.05\pm0.04)$\,PE/keV.  The use of $^{83\mathrm{m}}$Kr in this type of particle detector is a relatively new procedure, and is described in detail in \cite{Kastens:2009pa,Manalaysay:2009yq}.

For most of the measurements, the three grid electrodes in the LXe are grounded, along with the PMT photocathodes, ensuring that there are no electric fields within the LXe volume.  However, for three of the central scattering angles studied, data are also collected while applying high voltage (HV) to these electrodes, in order to measure the light yield's field dependence.  For this purpose, the cathode, gate, and anode grids are biased at $-$265\,V, $-$530\,V, and $-$2120\,V, respectively.  Due to effects of field leakage through the grid wires, the field in the main LXe volume (between the cathode and the gate, where the $\gamma$-ray beam is centered) is less than what would be naively determined by treating the electrodes as infinite solid plates.  In order to study the electric fields in all regions of the detector, its geometry and electrode voltages are simulated with the \textsc{comsol} Multiphysics\textsuperscript{\textregistered} software (version 4.3) \cite{comsolRef}.  These simulations show that the volume-averaged field strength in the main LXe region is $|\vec{\mathbf{E}}| = (450\pm8)$\,V/cm \cite{HrvojeBachelorThesis}.

\section{Monte-Carlo simulations}
\label{sec:mc}

Given the small scattering angles involved in this study, a detailed Monte Carlo (MC) simulation of the setup is essential to understand its systematic uncertainties.  All detector components are reproduced in detail with the GEANT4.9.3.p02 particle transport simulation code \cite{Agostinelli:2002hh,Allison:2006ve}.  The simulations additionally utilize a number of specialized physics packages that are tailored to low-energy electromagnetic processes, such as \texttt{G4LowEnergyCompton} \cite{LongoG4:2008}, which take into account modifications to the Compton process by the binding and kinetic energies of the electrons.

\begin{figure*}[htp!]
	\begin{center}
        \includegraphics[width=0.85\textwidth]{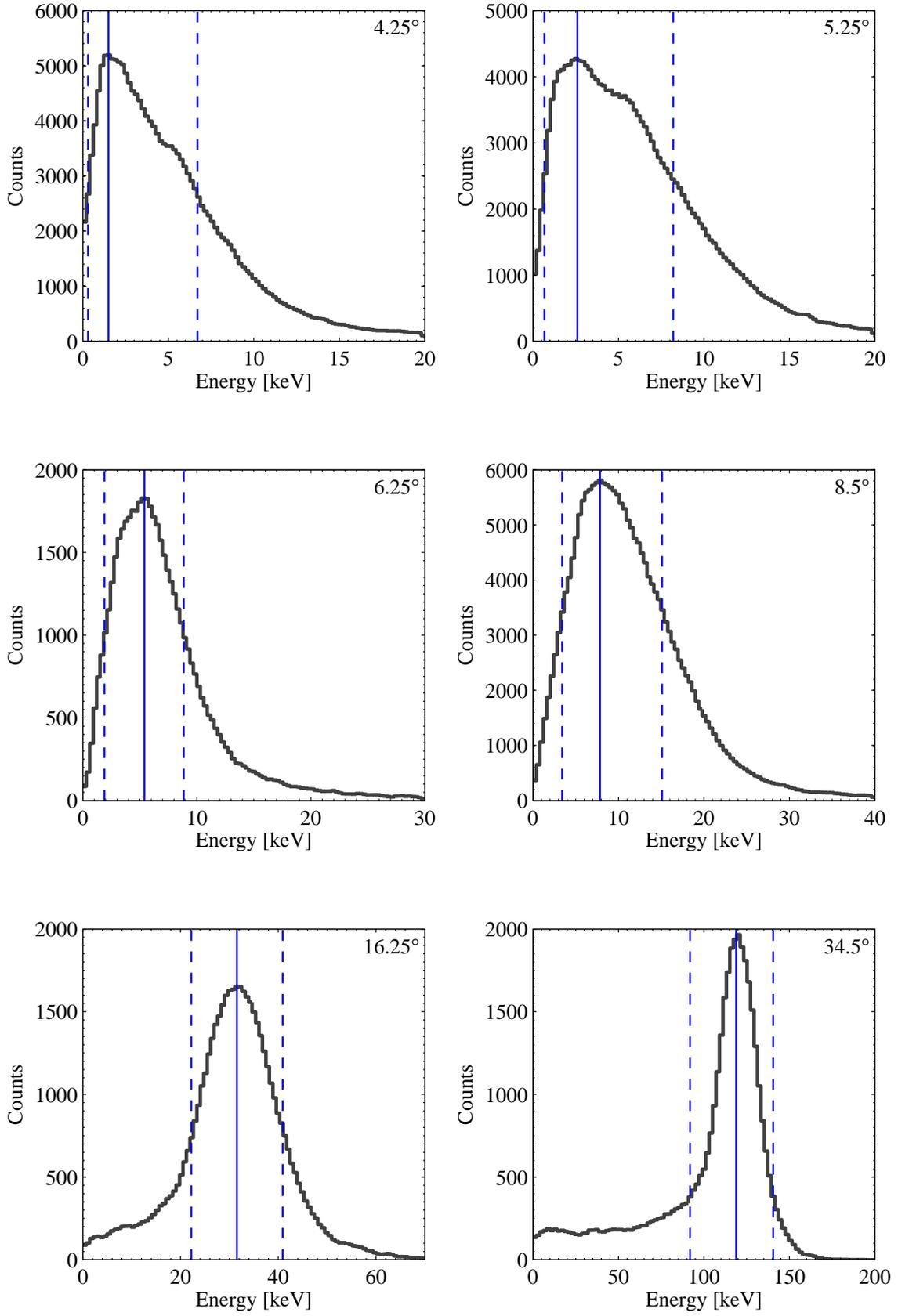} 
        \caption{Monte-Carlo distributions of raw energy deposition of Compton electrons (i.e.~no convolution or energy thresholds applied) for all central scattering angles.  Solid-blue lines indicate the peak position taken as the central energy, and the the blue-dashed lines enclose 68.3\% of the distribution, taken as the 1\,$\sigma$ spread in the energy deposition.  The feature in the spectra of the top-two plots at $\sim$5\,keV is due to Xe's $L$-shells whose binding energies are at 5.45\,keV, 5.10\,keV, and 4.78\,keV \cite{x-rayTransitionDB:2013}.}
        \label{fig:raw_spectra}
    \end{center}
\end{figure*}
Since the detector components subtend angles that are a non-negligible fraction of the central scattering angles, no data set features a perfectly monoenergetic energy deposition, but instead sees a range of energies.  The distribution of these energies must be understood if the underlying energy-dependent scintillation light yield is to be determined.  Figure \ref{fig:raw_spectra} shows the distribution of raw energy deposition for all scattering angles.  That is, no convolution or energy threshold has been applied, and these histograms represent the true distribution of recoil energies.  Two important features of these distributions are apparent.  First, the distributions are not symmetric, and instead feature an extended high-energy ``tail''.  Second, the peak of the distributions are shifted relative to the energy values predicted by Eq.~(\ref{eq:compt}).  These effects are a result of the fact that, although the distribution of scattering \emph{angles} is nearly symmetric, the recoil energy becomes quadratic in $\theta$ for small $\theta$ (that is, $E_{\mathrm{er}}\approx E_{\gamma}^2\theta^2/2m_{\mathrm{e}}c^2$).  Uniform intervals of energy, $\Delta E_{\mathrm{er}}$, therefore span intervals of $\theta$ that scale as $\theta^{-1}$, and hence smaller angles contribute more to the energy distribution than larger angles.  We take the central energy to be the peak position of the raw MC $E_{\mathrm{er}}$ distribution, and the 1\,$\sigma$ range given by a region covering 68.3\% of the total spectrum, shown in Table \ref{tab:LY_results}.  These bounds are chosen in a way such that the differential rate is equal at both boundaries (i.e.~the dashed-blue lines in Figure \ref{fig:raw_spectra} cross the black histogram at the same vertical position).

Events in which the $\gamma$ ray scatters multiple times in the LXe, or in detector materials other than active LXe or NaI, represent background populations that can be well understood with the MC.  The contributions from these backgrounds are small, since the small size of the LXe cell makes the former (``multiple scatters'') unlikely, and the latter (``materials scatters'') are reduced by considering only events within the full absorption peak in the NaI.  The simulations indicate that multiple scatters contribute on average 1.6\% to the total signal, and materials scatters on average 5.8\%.  

The results of the MC simulations are compared to data via the likelihood function (see Section \ref{sec:results}) in order to extract the light-yield values.  In an effort to account for potential misalignment of experimental components, central scattering angles are simulated additionally with $\theta\pm0.125^{\circ}$, and the range of resulting light yield values is taken as the geometrical systematic uncertainty.

\section{Analysis and Results}
\label{sec:results}
\subsection{Data selection}
\label{sec:results:data_selection}

\begin{figure}[htp!]
	\begin{center}
        \includegraphics[width=0.5\textwidth]{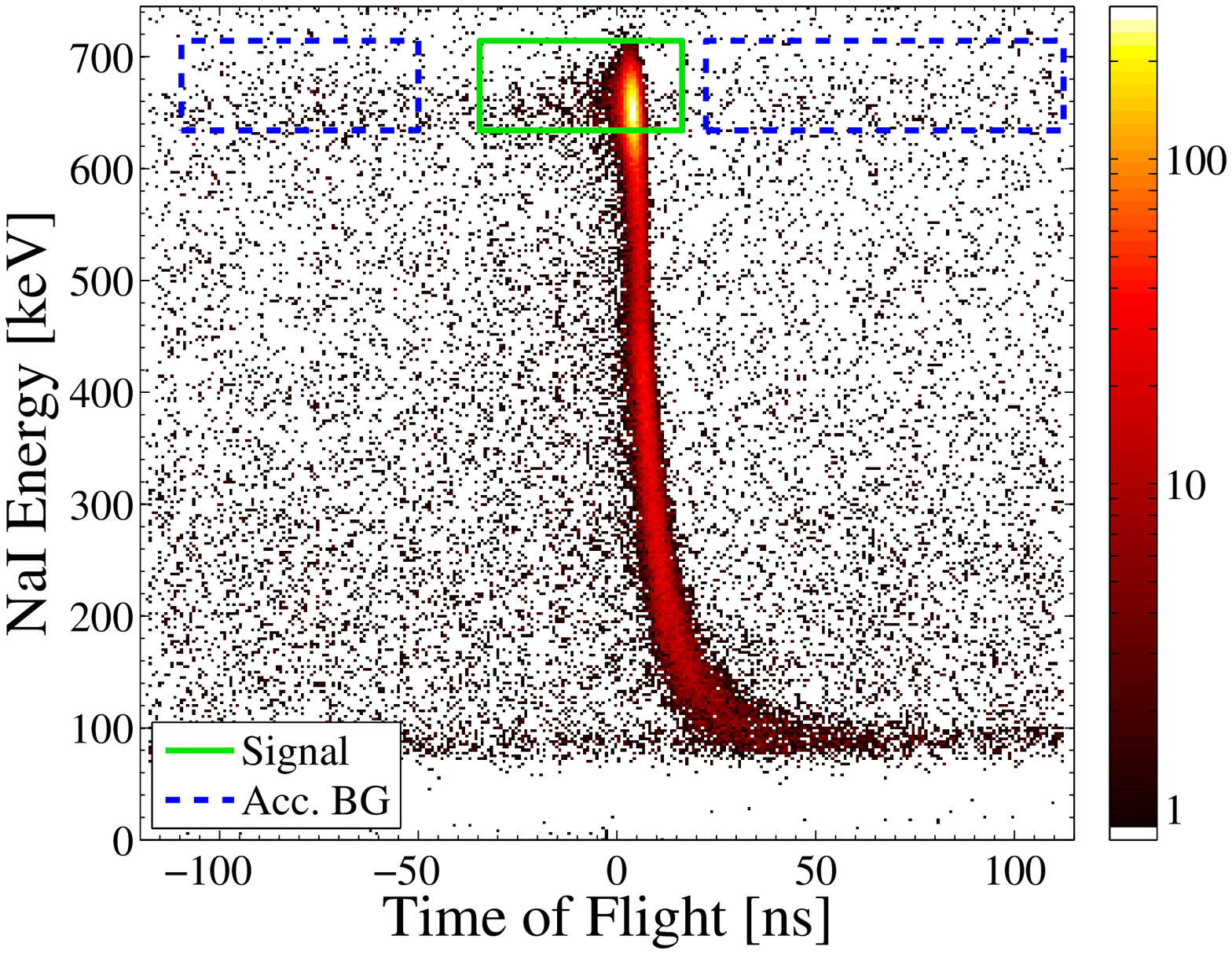} 
        \caption{(color online) Energy deposited in the NaI scintillator versus the time of flight of the $\gamma$ rays for the data with a central scattering angle of 8.5$^{\circ}$ ($E_{\mathrm{er}} = 7.8^{+7.3}_{-4.4}$).  The green (solid) box indicates the selection cuts used for the signal, blue (dashed) boxes are the sidebands used to estimate the background from accidental coincidences.}
        \label{fig:NaIvToF}
    \end{center}
\end{figure}
As mentioned in Section \ref{sec:methods}, four signals are recorded for each trigger: the PMT coupled to the NaI crystal, the two PMTs immersed in the LXe, and the output of the TAC.  Data quality cuts are performed to remove events that show significant shift or fluctuations in the baseline of the digitized trace.  The main data selection is performed based upon the NaI and TAC signals.  The two-dimensional distribution of these parameters, for the central scattering angle of 8.5$^{\circ}$, is shown in Figure \ref{fig:NaIvToF}.  The peak along the horizontal axis at roughly 4\,ns corresponds to the Time of Flight (ToF) of the $\gamma$ ray between the LXe and NaI.  The Compton continuum tends towards higher TAC values for lower energy depositions; this `walk effect' is a result of the fact that the start and stop signals controlling the TAC are generated by a leading-edge discriminator with a threshold that is given by an absolute voltage amplitude.  Therefore, a small signal will trigger the discriminator at a later time than a larger---but otherwise identically shaped---signal.  Though this feature could be removed by the use of a constant-fraction discriminator, it is inconsequential to the analysis because only events falling within the range [$-$1\,$\sigma$,+3\,$\sigma$] of the $^{137}$Cs full-absorption peak, located at (662\,keV-$E_{\mathrm{er}}$), are considered.  This asymmetric window is chosen in order to minimize the contribution from $\gamma$ rays that may undergo a small-angle Compton scatter along the transit from the LXe to the NaI.  A similar effect occurs with LXe trigger signals, but is significantly less pronounced than the NaI due to LXe's much faster scintillation pulse rise time.  A more significant effect on the ToF measurement for small LXe scintillation signals occurs due to the $\mathcal{O}$(10\,ns) scintillation emission timescale \cite{Hitachi:1983PSD}; for scintillation signals smaller than $\sim$10\,PE the arrival of the first PE can be delayed, resulting in a corresponding tail of the ToF distribution.  For this purpose, asymmetric ToF bounds are chosen in order to collect all emitted LXe photons.  

The green box in Figure \ref{fig:NaIvToF} indicates the bounds of the event selection cut.  Identical peak-selection cuts are applied to the MC events.  Though the distribution of events in ToF is highly peaked, there exists a population of events that is uniform in ToF, resulting from accidental coincidences between LXe and NaI signals.  The contribution of this background is estimated from the ToF sidebands enclosed by the dashed-blue boxes in Figure~\ref{fig:NaIvToF}.  The spectrum of LXe scintillation of the events in these sidebands shows no dependence on the ToF, and therefore the sideband boundaries are chosen to maximize the statistics of the estimate.  

A potential additional background that is coincident with the NaI signal arises from an observed effect whereby one of the PMTs gives a signal in response to a particle scatter in the PMT alone, likely the dynode chain.  This effect has been confirmed by placing one of the PMTs used here in a black box without any scintillator and comparing the dark-current spectrum with and without a $^{137}$Cs source in the vicinity.  A PMT signal of this sort is unconnected with any light emission, and therefore does not produce signals coincident in the two LXe PMTs.  In order to reduce this background, the $^{137}$Cs collimator is placed as close as possible to the LXe cryostat, resulting in a small spot size.  Since this background is absent from signals that are coincident in both LXe PMTs, the data from each scattering angle are each separately fit with the hardware-imposed N$\geq$1 coincidence requirement, and additionally while imposing an N$\geq$2 requirement in software.  The spectra obtained when requiring this more stringent selection requirement are corrected with a simulated efficiency curve, and the discrepancy between the fit results for N$\geq$1 and N$\geq$2 conditions is treated as a systematic uncertainty, shown in Table \ref{tab:LY_results}.

\subsection{Light-yield determination and results}
\label{sec:results:LY_det_results}
For each measured scattering angle, the light yield is determined by iteratively transforming the corresponding MC results into an expected scintillation distribution until the likelihood function is maximized.  The deposited energy, $E_i$, of each MC event $i$, is scaled by an energy-dependent light yield, $LY\!(E_i)$, to an expected scintillation signal, 
\begin{equation}
\label{eq:Si_Expect}
\langle S_i\rangle = E_i \times LY\!(E_i) , 
\end{equation}
in units of PE.  Because the raw energy distributions are not monoenergetic, $LY\!(E_i)$ should allow for a nonzero slope in the region of the peak energy (see Figure \ref{fig:raw_spectra}).  Outside the peak energy, $LY\!(E_i)$ should flatten, so that the tails of the distributions do not become scaled by unphysically large (or small) light yields.  Any generic sigmoid function can accomplish this purpose; we choose here the error function, such that $LY\!(E_i)\sim\mathrm{erf}[\ln(E_i)]$.  The argument of the error function is chosen to be logarithmic because the light yield is expected to change more at small energies than at large energies, and additionally the raw energy distributions are skewed with high-energy tails; a logarithmic energy scale is the simplest way to capture these features.  The $LY\!(E_i)$ function is shifted horizontally so that it is centered at the peak of the energy distribution, $E_c$, given a vertical offset, $LY_0$, and allowed to vary roughly within the range $[E_-,E_+]$ (the 1$\sigma$ bounds of the energy distribution).  With these adjustments, we take the functional form of the light yield to be
\begin{equation}
\label{eq:LY_enDep}
LY\!(E_i) = %LY_0+m\,E_c\,\frac{\sqrt{\pi}}{2}\,\mathrm{log}\!\left(\frac{E_+}{E_c}\right)\mathrm{erf}\!\left[\frac{\mathrm{log}(E_i/E_c)}{\mathrm{log}(E_+/E_c)}\right] ,
LY_0+\frac{m\,E_c\,\sqrt{\pi}}{2}\,\mathrm{ln}\!\left(\frac{E_+}{E_c}\right)\mathrm{erf}\!\left[\frac{\mathrm{ln}(E_i/E_c)}{\mathrm{ln}(E_+/E_c)}\right] ,
\end{equation}
where $E_c$ and $E_+$ are fixed constants for each scattering angle, and $m$ and $LY_0$ are free parameters.  The prefactors and argument factors are chosen so that this generic function has the convenient property that when $E_i=E_c$, both the function and its first derivative take on the simple forms $LY\!(E_c)=LY_0$ and $LY'(E_c)=m$.  The 1$\sigma$ lower bound on the energy distribution, $E_-$, does not explicitly appear in Eq.~(\ref{eq:LY_enDep}), but is implicitly respected because the energy distributions below 10$^{\circ}$ in Figure \ref{fig:raw_spectra} are nearly symmetric in log space.

The probability to obtain an integer number of PE, $j$, for a given MC event is found by assuming that the $\langle S_i\rangle$ from Eq.~(\ref{eq:Si_Expect}) each define the mean of a Poisson distribution,
\begin{equation}
\label{eq:S_ij}
S_{ij} = \frac{\langle S_i\rangle^j \,e^{-\langle S_i\rangle}}{j!} ,
%S_{ij} = \frac{\langle S_i\rangle^j \,\mathrm{exp}(-\langle S_i\rangle)}{j!} ,
\end{equation}
where $S_{ij}$ is the probability to obtain $j$ photoelectrons from the $i^{\mathrm{\,th}}$ MC event.  The expected number of events with $j$ photoelectrons for the entire MC run, $S_j$, is then given by summing the $S_{ij}$ over all MC events, $S_j = \sum_i S_{ij}$.

\begin{figure*}[hpt!]
	\begin{center}
        \includegraphics[width=0.98\textwidth]{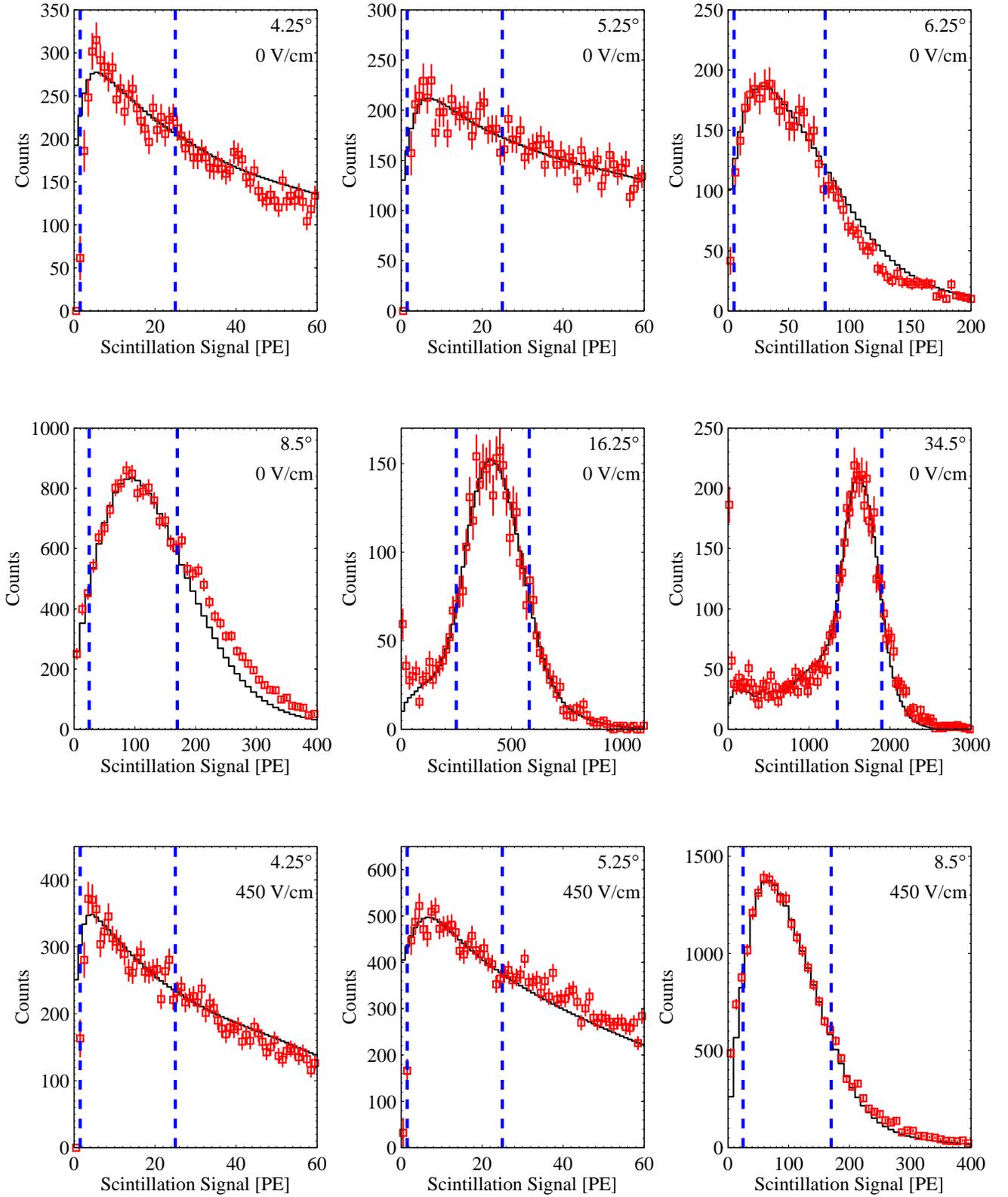} 
        \caption{(color online) Comparison between the scintillation spectra of real data (red boxes) and the simulated data using the best fit parameters (black) for all scattering angles.  Blue-dashed lines indicate the main fit range.  The discrepancy observed in the high-energy tails of the data obtained at 6.25$^{\circ}$ and 8.5$^{\circ}$ is absent when alternate fit ranges are used, with little effect on the reconstructed light yield (see text).}
        \label{fig:DatMC_spec}
    \end{center}
\end{figure*}

The spectrum $S_j$ is convolved with the measured single-PE response of the LXe PMTs to produce an expected measured signal (note that a measured signal can have fractional number of PE).  The top and bottom PMTs have single-PE resolutions ($\sigma/\mu$) of 54\% an 55\%, respectively.  An additional Gaussian convolution is applied to account for extra fluctuations that might arise from, for example, effects such as electron-ion recombination or position-dependent geometrical light collection.  In order to avoid making assumptions about the behavior of these effects, the functional form of the Gaussian convolution's width, $w$, is taken to be a power law in $j$ with a free exponent, $w(j)=w_1j^{w_2}+w_3$, which adds three free nuisance parameters to the fit ($w_1$, $w_2$, $w_3$).  Finally, this spectrum is re-binned to match the bin size of the real data, and then normalized according to the source activity, $A$.  This normalization is treated as a free parameter constrained with a Gaussian prior to account for the uncertainty in the activity.  This uncertainty is found by leaving $A$ completely free in initial fits to the 8.5$^{\circ}$, 16.25$^{\circ}$, and 34.5$^{\circ}$ datasets, and combining the resulting statistical uncertainties to form $A$'s prior.  The spectrum of accidentals background (estimated from the ToF sidebands) is added to the MC spectrum, forming the expected content of the $k^{\mathrm{th}}$ bin, $\nu_k$, which is implicitly a function of the six fit parameters, $\nu_k = \nu_k(LY_0,m,w_1,w_2,w_3,A)$.  The standard Poisson log-likelihood function, weighted with a Gaussian prior for the source activity, can be written as
\begin{equation}
\label{eq:likelihood}
\mathrm{ln}\mathcal{L} =\!\!\!\sum_{k\in\mathrm{f.r.}}\!\!(n_k\ln\nu_k - \nu_k) - \frac{A^2-2A\mu_{\!A}}{2\delta_{\!\!A}^{\,2}} ,
\end{equation}
where $n_k$ is the histogram of real data, $\mu_{\!A}$ and $\delta_{\!A}$ are the mean and width, respectively, of the Gaussian prior constraining the source activity.  The sum is carried out over bins within the fit range (f.r.), which is chosen to roughly cover the peak of the observed spectrum (because $LY\!(E)$ is featureless outside this peak region) and to avoid contributions from the long tails of the energy distributions.  In order to understand how the choice of the f.r.~affects the resulting $LY_0$, each f.r.~is later varied and the observed discrepancies in $LY_0$ are included as a systematic uncertainty (Section \ref{sec:results:systematics}).

\begin{table*}[htp!]
\begin{center}
%\begin{minipage}[b]{1.0\linewidth}
\caption{\label{tab:LY_results}Results of the light-yield measurements.  $\theta_{\mathrm{c}}$ is the central angle of the dataset; $E_{\mathrm{er}}$ is the central energy of the energy distribution; $\mathcal{R}_e$ is the zero-field central relative light yield value (relative to the scintillation emission at 32.1\,keV); $\sigma_{\mathrm{st}}$ is the statistical uncertainty;  $\sigma_{\mathrm{sys}}^{(1)}$ is the systematic uncertainty resulting from potential misalignment of experimental components; $\sigma_{\mathrm{sys}}^{(2)}$ is the systematic uncertainty associated with the choice of fit range; $\sigma_{\mathrm{sys}}^{(3)}$ is the systematic uncertainty associated with source activity; $\sigma_{\mathrm{sys}}^{(4)}$ indicates the discrepancy introduced between 1-fold and 2-fold coincidence requirements on the LXe PMTs; an additional systematic uncertainty of 1.5\% is applicable to all values in the third column, which arises from variations in results of weekly $^{57}$Co calibrations.  $q(450)$ is the scintillation quenching factor at an applied field of 450\,V/cm; the first uncertainties are statistical, the second systematic.} 
	\begin{tabular*}{0.8\textwidth}{@{\extracolsep{\fill}}c | c | c | c | c | c | c | c || c }
		\hline \hline
		\multirow{2}{*}{$\theta_{\mathrm{c}}$} & 
		\multirow{2}{*}{$E_{\mathrm{er}}$\,{\small(keV)}} & 
		\multirow{2}{*}{$\mathcal{R}_e$} & 
		\multirow{2}{*}{$\sigma_{\mathrm{st}}$} & 
		\multirow{2}{*}{$\sigma_{\mathrm{sys}}^{(1)}$} & 
		\multirow{2}{*}{$\sigma_{\mathrm{sys}}^{(2)}$} & 
		\multirow{2}{*}{$\sigma_{\mathrm{sys}}^{(3)}$} & 
		\multirow{2}{*}{$\sigma_{\mathrm{sys}}^{(4)}$} & 
		\multirow{2}{*}{$q(450)$} \\
		& & & & & & & &  \\
		\hline
		\multirow{2}{*}{4.25$^{\circ}$}  & 
			\multirow{2}{*}{1.5$^{+5.2}_{-1.2}$} & 
			\multirow{2}{*}{0.37} & 
			\multirow{2}{*}{$^{+0.20}_{-0.12}$} &
			\multirow{2}{*}{$^{+0.03}_{-0.04}$} & % ang
			\multirow{2}{*}{$\pm$0.03} & % f.r.
			\multirow{2}{*}{$\pm$0.02} & % sig_a
			\multirow{2}{*}{$\pm$0.14} & % n>=2
			\multirow{2}{*}{$0.64^{+0.45+0.09}_{-0.20-0.09}$} \\ % field
		& & & & & & & & \\
		5.25$^{\circ}$  & 
			2.6$^{+5.6}_{-1.9}$ & 
			0.52 & 
			$^{+0.10}_{-0.15}$ & 
			$^{+0.03}_{-0.03}$ & % ang
			$\pm$0.01 & % f.r.
			$\pm$0.06 & % sig_a
			$\pm$0.05 & % n>=2
			$0.77^{+0.42+0.02}_{-0.28-0.02}$ \\ % field
		\multirow{2}{*}{6.25$^{\circ}$}  & 
			\multirow{2}{*}{5.4$^{+3.5}_{-3.5}$} & 
			\multirow{2}{*}{0.57} & 
			\multirow{2}{*}{$^{+0.08}_{-0.15}$} & 
			\multirow{2}{*}{$^{+0.03}_{-0.02}$} & % ang
			\multirow{2}{*}{$\pm$0.04} & % f.r.
			\multirow{2}{*}{$\pm$0.01} & % sig_a
			\multirow{2}{*}{$\pm$0.03} & % n>=2
			\multirow{2}{*}{---} \\ % field
		& & & & & & & & \\
		8.50$^{\circ}$  & 
			7.8$^{+7.3}_{-4.4}$ & 
			0.82 & 
			$^{+0.03}_{-0.02}$ & 
			$^{+0.03}_{-0.03}$ & % ang
			$\pm$0.03 & % f.r.
			$\pm$0.04 & % sig_a
			$\pm$0.01 & % n>=2
			$0.74^{+0.03+0.12}_{-0.03-0.12}$ \\ % field
		\multirow{2}{*}{$^{83\mathrm{m}}$Kr} & 
			\multirow{2}{*}{9.4} & 
			\multirow{2}{*}{1.10} & 
			\multirow{2}{*}{$^{+004}_{-004}$} & 
			\multirow{2}{*}{---} & % ang
			\multirow{2}{*}{---} & % f.r.
			\multirow{2}{*}{---} & % sig_a
			\multirow{2}{*}{---} & % n>=2
			\multirow{2}{*}{$0.893^{+0.001+0.014}_{-0.001-0.014}$} \\ % field		
		& & & & & & & & \\
		16.25$^{\circ}$ & 
			31.6$^{+9.4}_{-9.4}$ & 
			0.96 & 
			$^{+0.01}_{-0.01}$ & 
			$^{+0.01}_{-0.02}$ & % ang
			$\pm$0.01 & % f.r.
			$\pm$0.01 & % sig_a
			$\pm$0.00 & % n>=2
			--- \\ % field
		\multirow{2}{*}{$^{83\mathrm{m}}$Kr} & 
			\multirow{2}{*}{32.1} & 
			\multirow{2}{*}{$\equiv 1$} & 
			\multirow{2}{*}{---} & 
			\multirow{2}{*}{---} & % ang
			\multirow{2}{*}{---} & % f.r.
			\multirow{2}{*}{---} & % sig_a
			\multirow{2}{*}{---} & % n>=2
			\multirow{2}{*}{$0.741^{+0.001+0.011}_{-0.001-0.011}$} \\ % field		
		& & & & & & & & \\
		34.50$^{\circ}$ & 
			118.9$^{+21.6}_{-27.0}$ & 
			0.959 & 
			$^{+0.005}_{-0.004}$ & 
			$^{+0.005}_{-0.006}$ & % ang
			$\pm$0.005 & % f.r.
			$\pm$0.008 & % sig_a
			$\pm$0.000 & % n>=2
			--- \\ % field
		\multirow{2}{*}{$^{57}$Co} & 
			\multirow{2}{*}{126.1} & 
			\multirow{2}{*}{0.97} & 
			\multirow{2}{*}{$^{+0.003}_{-0.003}$} & 
			\multirow{2}{*}{---} & % ang
			\multirow{2}{*}{---} & % f.r.
			\multirow{2}{*}{---} & % sig_a
			\multirow{2}{*}{---} & % n>=2
			\multirow{2}{*}{$0.593^{+0.003+0.009}_{-0.003-0.009}$} \\ % field		
		& & & & & & & & \\
		\hline \hline
	\end{tabular*}
%\end{minipage}
\end{center}
\end{table*}

The likelihood function (Eq.~(\ref{eq:likelihood})) is proportional to the Bayesian posterior probability density function (PDF). This PDF is sampled with the Metropolis-Hastings algorithm \cite{chib:95,Saha:02}, which is an iterative Markov-Chain process that is well suited for producing a random sample of data from a multidimensional PDF for which evaluation may be computationally intensive.  It has the additional useful property that, because the output is a set of data, marginalizing the posterior PDF over nuisance parameters ($m$, $w_i$, and $A$) is as simple as histogramming the desired parameter, $LY_0$.  The data collected at all scattering angles are shown in Figure \ref{fig:DatMC_spec}, along with the corresponding best-fit MC spectra.  The statistical uncertainties are given by the 1$\sigma$ contour of the marginal posterior PDF in $LY_0$.  The subtle \textit{L}-shell feature seen in the raw-energy spectra (Figure \ref{fig:raw_spectra}) at 4.25$^{\circ}$ and 5.25$^{\circ}$ is not visible in the scintillation data due to Poisson smearing.

\begin{figure}[htp!]
	\begin{center}
        \includegraphics[width=0.49\textwidth]{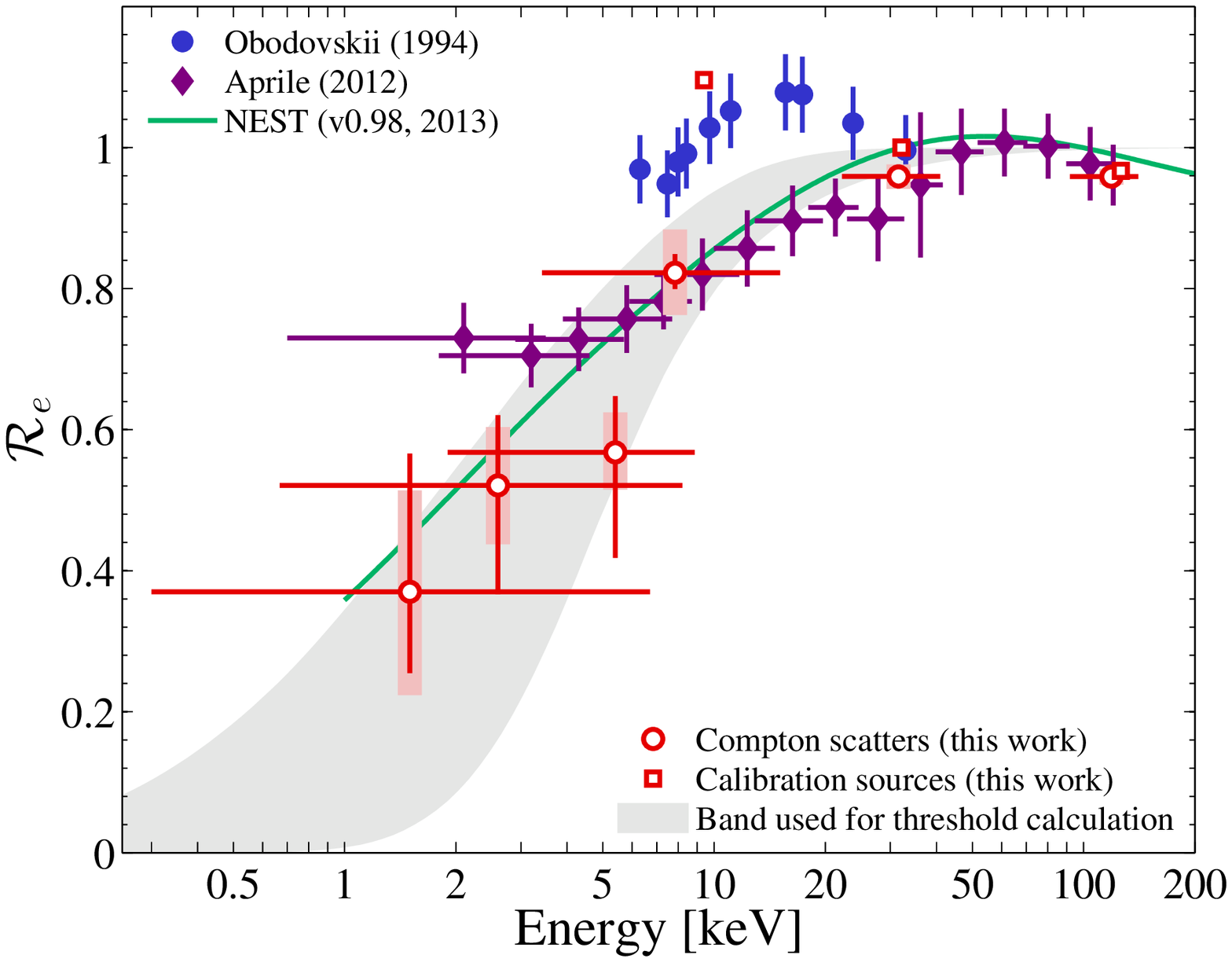} 
        \caption{(color online) Results of the light yield relative to that of the 32.1\,keV emission of $^{83\mathrm{m}}$Kr, $\mathcal{R}_e$.  The current work (red) shows statistical uncertainties as vertical lines, systematic uncertainties as light, shaded rectangles, and the 1\,$\sigma$ spread in the distribution of electron recoil energies as horizontal lines.  Also shown are the results from studies with X-rays \cite{obodovskii:1994} (blue), the recent Compton-scatter study by Aprile \textit{et al}.~\cite{Aprile:2012an} (purple) and the model prediction of NEST \cite{Szydagis:2011tk,NEST:web} (green).  The gray band indicates the 1$\sigma$ range of $\mathcal{R}_e$ models used to determine the energy thresholds of four recent LXe dark-matter searches.}
        \label{fig:LY_0field}
    \end{center}
\end{figure}

Following the procedure of \cite{Aprile:2012an}, we present $\mathcal{R}_e$, the LY results relative to the 32.1\,keV emission of $^{83\mathrm{m}}$Kr, shown in Table \ref{tab:LY_results} and Figure \ref{fig:LY_0field} (the reason for this normalization is discussed in Section \ref{sec:discussion:9keV_anomaly}).  Also shown are the corresponding $\mathcal{R}_e$ values from the $^{57}$Co and $^{83\mathrm{m}}$Kr sources, along with the results of Obodovskii and Ospanov using induced X-rays \cite{obodovskii:1994}, the recent measurement by Aprile \textit{et al.}~\cite{Aprile:2012an}, and the prediction from the Noble Element Simulation Technique (NEST) version 0.98 \cite{Szydagis:2011tk,NEST:web}, which is a constrained implementation of the Thomas-Imel electron-ion recombination model \cite{Thomas_j:1987}.  The values from Obodovskii and the curve from NEST have been vertically normalized such that their interpolated values at 32.1\,keV are unity.  The gray band indicates the 1$\sigma$ allowed LY models used in energy-threshold determinations, discussed in Section \ref{sec:discussion:dm_searches}.

\subsection{Systematic uncertainties}
\label{sec:results:systematics}
Five systematic uncertainties are considered, as shown in Table \ref{tab:LY_results}. 
\begin{itemize}
\item$\sigma^{(\mathrm{1})}$ results from potential misalignment of the experimental components (discussed in Section \ref{sec:mc}), and is studied by simulating central angles with a shift of $\pm0.125^{\circ}$ and determining the best $LY_0$ as before.  
\item$\sigma^{(\mathrm{2})}$, the uncertainty from the choice of fit ranges, is studied by modulating the chosen fit ranges at the level of 20\% and taking the observed differences in the best-fit $LY_0$ values.  The spectra from 6.25$^{\circ}$ and 8.5$^{\circ}$ in Figure \ref{fig:DatMC_spec} show a discrepancy between data and MC in the high-energy tail; this discrepancy is absent in the fit obtained with the alternate fit range, with little effect on the reconstructed light yield.  
\item$\sigma^{(\mathrm{3})}$ indicates the dependence of the best-fit $LY_0$ on the source-activity parameter, $A$.  This systematic is determined by using the covariance between $LY_0$ and $A$ from the fit to estimate how much $LY_0$ can vary within the allowed uncertainty on $A$, given by
\begin{equation}
\label{eq:A_slope_sys}
\sigma^{(\mathrm{3})} = \frac{\mathrm{cov}(LY_0,A)}{\sigma^2_A}\,\delta_{\!A},
\end{equation}
where $\delta_{\!A}$ is the uncertainty in the source activity (as in Eq.~(\ref{eq:likelihood})) and $\sigma^2_A$ is the variance of $A$ from the fit.  The factor $\mathrm{cov}(LY_0,A)/\sigma^2_A$ gives the slope of $LY_0$ versus $A$.  
\item$\sigma^{(\mathrm{4})}$ quantifies the uncertainty associated with the choice of the PMT coincidence requirement.  An $N=2$ coincidence requirement on the two LXe PMTs is separately imposed, correcting the resulting scintillation spectrum by a simulated coincidence efficiency curve, and performing the fits again for $LY_0$.  
\item$\sigma^{(\mathrm{5})}$ is a 1.5\% relative systematic from fluctuations in the PMT gains and weekly $^{57}$Co calibrations.  
\end{itemize}
These systematic uncertainties are combined in quadrature to form the systematic error bars in Figure \ref{fig:LY_0field}, and the first four are shown in Table \ref{tab:LY_results}.  In the lowest energy, the dominating systematic is $\sigma^{(4)}$ with a contribution of 38\%; this systematic rapidly decreases to 1\% by 8.5$^{\circ}$ and zero beyond.

\subsection{Field dependence}
\label{sec:results:field}
\begin{figure}[htp!]
	\begin{center}
        \includegraphics[width=0.49\textwidth]{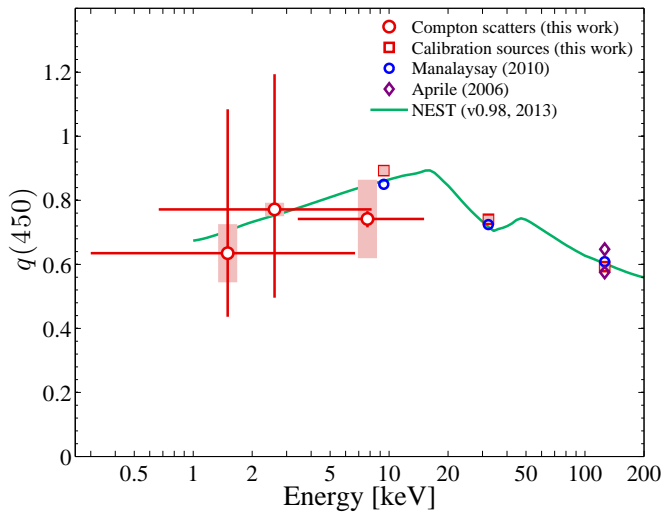} 
        \caption{(color online) The quenching of the scintillation signal with an applied electric field of 450\,V/cm.  Vertical lines represent statistical uncertainties, grey bars represent systematic uncertainties, and horizontal lines are the 1\,$\sigma$ spread in the distribution of electron recoil energies.  Also show are the parameterized predictions from \cite{Manalaysay:2009yq} (blue circles) and $^{57}$Co field quenching \cite{Aprilenc:2006} (purple diamonds) at 400\,V/cm and 500\,V/cm.  The prediction of the NEST model \cite{Szydagis:2011tk,NEST:web} for quenching at 450\,V/cm is indicated by the green curve.}
        \label{fig:quenching}
    \end{center}
\end{figure}
The previous results all pertain to the light yield of LXe with no applied electric fields.  As mentioned in Section \ref{sec:methods}, data were also collected with an applied field of 450\,V/cm for a subset of scattering angles in order to study the scintillation quenching of LXe at the lowest energies.  The data collected with this field are fit using the same procedure as before, resulting in a set of posterior PDFs for the light yield.  The last row of Figure \ref{fig:DatMC_spec} shows the measured and best-fit spectra of the three scattering angles collected.  These PDFs are convolved with their corresponding zero-field light yield PDFs to obtain posterior PDFs of their ratio, known as the field-quenching value, $q(450)$, shown in Table \ref{tab:LY_results}.  For each scattering angle with applied field, the 450\,V/cm data and the zero-field data were taken consecutively.  Therefore, any potential misalignment of experimental components will be unrelated to the applied field.  The resulting scintillation quenching values, along with those simultaneously obtained for $^{57}$Co and $^{83\mathrm{m}}$Kr, are shown in Figure~\ref{fig:quenching}.  Also shown is the predicted scintillation quenching of the NEST model.

\section{Discussion}
\label{sec:discussion}
\subsection{Comparison of results}
\label{sec:discussion:results_comparison}
The results presented here represent the first observation of LXe scintillation light from electronic recoils down to 1.5\,keV, and additionally measure the behavior of this scintillation emission under the application of a static electric field.  The general behavior---that of reduced LY for decreasing energies---is predicted by a number of methods (see \cite{Szydagis:2011tk} and references therein), and is understood as being due to reduced electron-ion recombination.  Below 10\,keV, the data show no significant energy dependence on the strength of field quenching, but support an average value of $q(450) = 0.74\pm0.11$.  For the NEST prediction of this quantity shown in Figure \ref{fig:quenching}, the horizontal scale indicates the energy of the primary $\gamma$ ray (not electronic-recoil energy), and is therefore in principle distinct from Compton scatters.  The feature in the NEST curve between $\sim$15\,keV and $\sim$50\,keV is an indirect result of photoabsorption on \textit{K}-shell electrons, and would be absent for Compton scatters of this energy.  However, the distinction between Compton scatters and photoabsorptions disappears at low energies \cite{Szydagis:2011tk,Szydagis:priv13}, where the recombination probability becomes independent of stopping power, and instead depends only on the total number of charges produced.  It is therefore an applicable prediction of our results in this energy regime.

It is interesting to note that the data obtained from X-rays \cite{obodovskii:1994} show an increased light yield at 7.84\,keV compared with the data obtained here from Compton scatters, when normalizing their interpolated value at 32.1\,keV.  The photoabsorption process that the X-rays undergo favors inner-shell electrons (when accessible) \cite{Knoll:00}, which means that the recoiling electrons can have significantly less energy than the incoming photons because they must overcome large binding energies.  On the other hand, Compton scattering on inner-shell electrons is suppressed for scattering angles below $\sim$60$^{\circ}$ \cite{Talukdar197137}.  Therefore, the two results actually probe LXe's response at slightly different electron energies.  In principle, the axioelectric effect, which has been induced as a possible explanation of the observed DAMA annual modulation signal, would be similar to the photoelectric effect.  However there is of course an overlap of effects, since low-energy Compton scatters do also probe inner-shell electrons, as can be seen by the $L$-shell feature in Figure \ref{fig:raw_spectra}.

The data reported by Aprile \textit{et al}.~\cite{Aprile:2012an} show good agreement with the present results above $\sim$10\,keV, but show a separation below this energy.  Considering both statistical and systematic uncertainties gives a maximum discrepancy of 1.7$\sigma$ at $\sim$5\,keV and 1.4$\sigma$ at $\sim$1.5\,keV.  

\subsection{The 9.4\,keV anomaly}
\label{sec:discussion:9keV_anomaly}
The discrepancy seen in the LY of the 9.4\,keV emission from $^{83\mathrm{m}}$Kr deserves attention.  The energy of this decay is carried mostly by internal conversion electrons emitted from the inner shell \cite{DeMin:1995tk}, however, this data point is inconsistent also with the X-ray data, for which the process should in principle be similar.  One notable characteristic of the 9.4\,keV emission is that it quickly follows the 32.1\,keV emission of the same nucleus, with a half-life of 154.4\,ns \cite{Ekstroem}.  It was pointed out by \cite{Plante:priv12} that the 32.1\,keV emission could leave behind a cloud of electron-ion pairs, close to the mother nucleus, that fail to recombine.  The electrons (ions) produced by the 9.4\,keV emission could then potentially have an additional supply of left-over ions (electrons) with which to recombine, producing more scintillation photons than would be observed normally.

\begin{figure}[htp!]
	\begin{center}
        \includegraphics[width=0.49\textwidth]{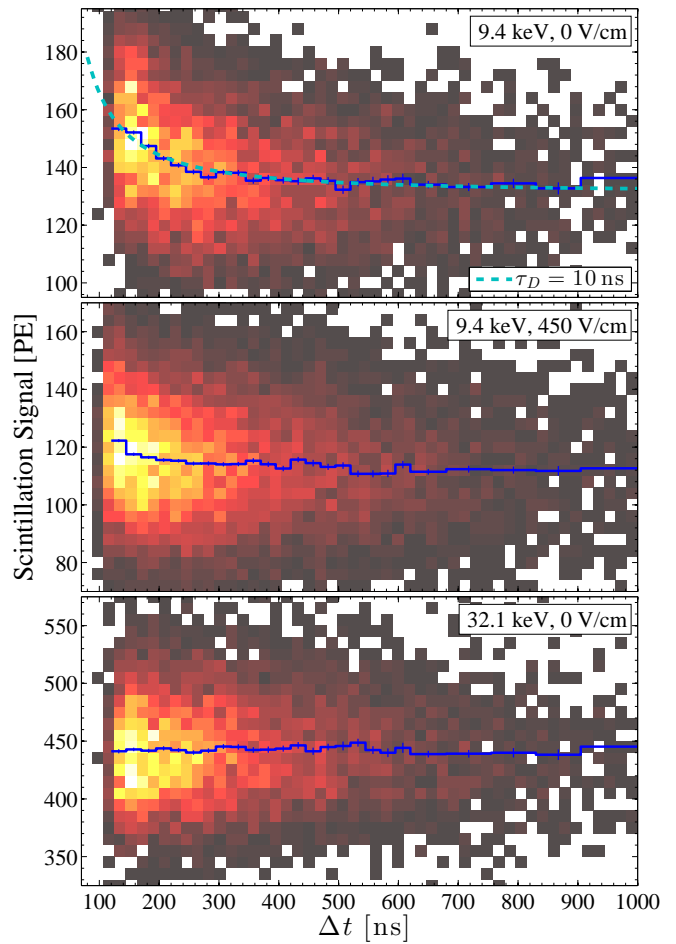} 
        \caption{(color online) The distribution of scintillation signals from the zero-field 9.4\,keV emission of $^{83\mathrm{m}}$Kr (top) as a function of the time since the same nucleus's 32.1\,keV transition, $\Delta t$.  The dashed line indicates the expected time dependence for diffusion of electrons from a 32.1\,keV interaction.  Also shown are the corresponding distributions for 9.4\,keV at 450\,V/cm (middle) and 32.1\,keV at zero field (bottom).  In all panels, the blue histograms indicate the mean of the scintillation distributions in progressive time slices.}
        \label{fig:Kr9_vDT}
    \end{center}
\end{figure}

We test this possibility by investigating the time dependence of the 9.4\,keV LY, since the scintillation enhancement should disappear as these charges diffuse away over time.  Figure \ref{fig:Kr9_vDT} (\emph{top}) shows the size of the 9.4\,keV scintillation signal at zero field as a function of delay between 32.1\,keV and 9.4\,keV emissions, $\Delta t$.  A clear rise in scintillation signal is seen for delay times $\lesssim$400\,ns.  Although making a precise prediction for the diffusion rate of the electrons and ions left over from the 32.1\,keV emission is difficult, a simple estimate can provide a useful expectation.  For an initial population of particles that is distributed according to a spherically symmetric Gaussian, it is straightforward to show that the central number density, $n_c$, evolves with time according to 
\begin{equation}
	\label{eq:GaussDiffusion}
	n_c \propto \left(2t+\frac{a^2}{D}\right)^{-3/2} ,
\end{equation}
where $a$ is a length scale characterizing the initial size of the distribution, and $D$ is the diffusion coefficient of the particles in the medium.  The quantity $a^2/D$ has dimensions of time and can be associated with a \textit{diffusion timescale}, $\tau_D$, given by
\begin{equation}
	\label{diffusionTime}
	\tau_{D} = \frac{a^2}{D} .
\end{equation}
The typical range of a 30\,keV electron in LXe is on the order of $\sim$10\,$\mu$m \cite{ESTAR} and the diffusion coefficient of electrons in LXe is $D_-\sim60$\,cm$^2$/s \cite{Doke198287}.  The ionic mobility of Xe$^+$ is given in the literature as $\mu_+\approx4\times10^{-3}$\,cm$^2$/Vs \cite{Hilt:1994a}, which can be connected to its diffusion coefficient by the Nernst-Einstein relation \cite{Hilt:1994b},
\begin{equation}
D_+ = \frac{kT}{e}\mu_+ ,
\end{equation}
giving $D_+\sim6\times10^{-5}$\,cm$^2$/s at $T=180$\,K.  With these values for $a$ and $D_{\pm}$, we can estimate the diffusion timescales as
\begin{align}
\label{eq:diffRates}
\tau_{D-}&~~=~~\mathcal{O}(10 \mathrm{\,ns})& \quad (&e^-) \\
\tau_{D+}&~~=~~\mathcal{O}(10 \mathrm{\,ms})& \quad (&\mathrm{Xe}^+) .
\end{align}
To illustrate how this compares to the observed behavior, Eq.~(\ref{eq:GaussDiffusion}) is plotted in Figure \ref{fig:Kr9_vDT} (\textit{top}) for $\tau_D=10$\,ns (dashed-cyan curve).  The same curve using $\tau_D=10$\,ms appears simply as a straight line on this scale, which is not surprising---given the 154\,ns half-life of the 9.4\,keV state, it would be unlikely to see a change in its LY due to Xe$^+$ diffusion.  However, these leftover ions could still enhance the scintillation signal of the 9.4\,keV transition, albeit with no observable time dependence.  The electron diffusion timescale estimated here is consistent with the observed time dependence.  It is therefore reasonable to conclude that the anomalous LY of the 9.4\,keV emission from $^{83\mathrm{m}}$Kr is due to left-over charges from the preceding 32.1\,keV emission.

An applied electric field provides an extra means (in addition to diffusion) by which electrons can leave the vicinity of the decaying nucleus.  The timescale at which this occurs can be estimated by the time required for the charges to traverse the affected region, given by
\begin{equation}
\label{eq:driftRates}
\tau_{\mu-} \sim \frac{a}{|\vec{\mathbf{E}}|\mu_-} = \frac{a}{v_d} = 5 \mathrm{\,ns},
\end{equation}
where $\mu_-$ is the electron mobility, $|\vec{\mathbf{E}}|$ is the electric field strength, and $v_d \approx 2$\,mm/$\mu$s is the drift velocity of electrons \cite{Miller:1968zz}.  Given that this process is of similar order to the electron diffusion process, one expects the scintillation signal at $|\vec{\mathbf{E}}|=450$\,V/cm to show less of an effect from electron diffusion.  This time dependence is shown in Figure \ref{fig:Kr9_vDT} (\textit{middle}), and exhibits a reduced scintillation increase at low $\Delta t$.

Though the average zero-field LY of these data gives $\mathcal{R}_e=1.10$ as quoted in Table \ref{tab:LY_results}, this value depends on the range of $\Delta t$ considered.  Since the characteristic PMT output from a $^{83\mathrm{m}}$Kr decay is a trace containing two scintillation pulses, the pulse-finding algorithm used must correctly identify this feature.  However, this can be difficult for small $\Delta t$, when the pulses begin to overlap.  The efficiency for the pulse-finding algorithm implemented here to correctly identify such double-pulse events is unity for $\Delta t\gtrsim150$\,ns, but this would not necessarily be true for other detectors utilizing a different analog bandwidth, data acquisition system, and data processing techniques.  For these reasons, $^{83\mathrm{m}}$Kr's 9.4\,keV transition is not very well suited as a ``standard candle'' calibration source in the way that $^{57}$Co is often implemented.  In principle, the 9.4\,keV transition could be used for standardized calibrations if only events in which $\Delta t>400$\,ns are considered.  However, although the results presented here show no significant time dependence above this value, the results in \cite{Aprile:2012an} do exhibit a continued decrease until roughly 1\,$\mu$s, and therefore one cannot predict how significant this observed time dependence might be in various other systems.  On the other hand, the 32.1\,keV transition shows no dependence on $\Delta t$, as expected, which we show here in Figure \ref{fig:Kr9_vDT} (\textit{bottom}).  We therefore conclude, in agreement with \cite{Aprile:2012an}, that the 32.1\,keV transition of $^{83\mathrm{m}}$Kr provides a good calibration source by which to compare the scintillation response of various LXe detectors.

\subsection{Impact on dark matter searches}
\label{sec:discussion:dm_searches}
The primary motivation for the present study is to learn whether existing LXe dark matter search results have energy thresholds that are low enough to probe a 2-5\,keV electronic-recoil peak, as potentially observed in the DAMA/LIBRA experiment.  To illustrate how our results can address this question, we consider four existing LXe dark-matter-search results: ZEPLIN-III \cite{Akimov:2011tj}, XENON10 \cite{Angle:2007uj}, XENON100 \cite{Aprile:2012nq}, and XMASS \cite{Abe:2012az}.  We wish to determine their electronic-recoil energy thresholds based on their quoted scintillation thresholds.  Given an electronic recoil that deposits energy $E_{\mathrm{er}}$, the average scintillation signal, $S1$, in units of PE will be given by
\begin{equation}
\label{eq:E_threshold}
S1 = E_{\mathrm{er}}\times f_{\mathrm{Co}}(E_{\mathrm{er}})\times LY_{\mathrm{Co}}\times\frac{q(|\vec{\mathbf{E}}|)}{q_{\mathrm{Co}}}
\end{equation}
where $f_{\mathrm{Co}}(E_{\mathrm{er}})$ is the ratio of the zero-field LY at $E_{\mathrm{er}}$ to that from $^{57}$Co, $\vec{\mathbf{E}}$ is the applied field, $LY_{\mathrm{Co}}$ is the $^{57}$Co light yield (in PE/keV) at $\vec{\mathbf{E}}$, $q_{\mathrm{Co}}$ is the scintillation quenching of $^{57}$Co at $\vec{\mathbf{E}}$, and $q(|\vec{\mathbf{E}}|)$ is the scintillation quenching at energy $E_{\mathrm{er}}$ and field $\vec{\mathbf{E}}$.  To determine an energy threshold, Eq.~(\ref{eq:E_threshold}) is inverted and evaluated for the quoted scintillation threshold, $S1_{\mathrm{thr}}$.

\begin{figure}[htp!]
	\begin{center}
        \includegraphics[width=0.49\textwidth]{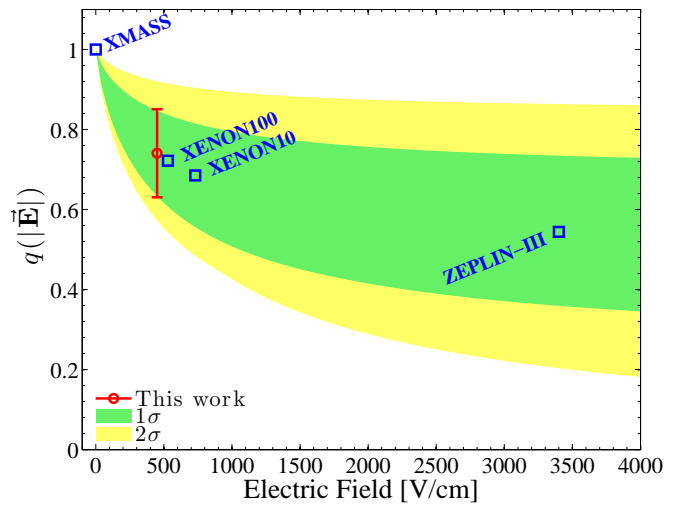} 
        \caption{(color online) The 1$\sigma$ and 2$\sigma$ bands of the scintillation field quenching below $\sim$10\,keV, $q(|\vec{\mathbf{E}}|)$, used in the determination of $E_{\mathrm{thr}}$.  Also indicated are the fields utilized by the four dark-matter experiments considered in the text.  The red circle indicates the measured $q(450)$, averaged from data obtained at 4.25$^{\circ}$, 5.25$^{\circ}$, and 8.5$^{\circ}$.}
        \label{fig:q_vs_field}
    \end{center}
\end{figure}

All four experiments utilize different applied electric fields, and therefore we must extrapolate the field-quenching results reported here to the appropriate values.  This extrapolation introduces an uncertainty in the calculated energy threshold that ranges from nonexistent in XMASS ($|\vec{\mathbf{E}}|=0$) to negligible in XENON100 ($|\vec{\mathbf{E}}|=530$\,V/cm) to considerable in ZEPLIN-III ($|\vec{\mathbf{E}}|=3400$\,V/cm).  In order to do so, we use an empirical parameterization of the field quenching inspired by the Thomas-Imel electron-ion recombination model \cite{Thomas_j:1987}, which was applied to LXe's $^{83\mathrm{m}}$Kr response in \cite{Manalaysay:2009yq} as
\begin{equation}
\label{eq:field_quench}
q(|\vec{\mathbf{E}}|) = a_1a_2|\vec{\mathbf{E}}|\,\ln\!\!\left(1+\frac{1}{a_2|\vec{\mathbf{E}}|}\right)+1, 
\end{equation}
where $a_1$ and $a_2$ are free parameters, with $a_1$ describing the overall strength of the field quenching, and $a_2$ describing the field dependence of this quenching.  In \cite{Manalaysay:2009yq}, it was found that the energy dependence of $a_2$ is much less significant than that of $a_1$.  Therefore, we use here $a_2 = (8.3\pm1.7)\times10^{-4}$\,cm/V, which is the average of the values for 9.4\,keV and 32.1\,keV reported in \cite{Manalaysay:2009yq}.  The value of $a_1$ is chosen so that the $q(|\vec{\mathbf{E}}|)$ function is consistent with our $q(450)$ value, which is taken from the average of data collected at 4.25$^{\circ}$, 5.25$^{\circ}$, and 8.5$^{\circ}$ (indicated by the red circle in Figure \ref{fig:q_vs_field}).  Combining the uncertainties in $a_2$ and $q(450)$ produces the bands show in Figure \ref{fig:q_vs_field}, which is taken to represent the energy-averaged field quenching below $\sim$10\,keV.

Also needed in the determination of the electronic-recoil energy threshold, $E_{\mathrm{thr}}$, is a model of $f_{\mathrm{Co}}(E_{\mathrm{er}})$.  For this, a range of models are taken that fit our $\mathcal{R}_e$ data, and whose 1$\sigma$ span is indicated by the gray band in Figure \ref{fig:LY_0field}.  The uncertainties on these three parameters ($a_2$, $q(450)$, and $f_{\mathrm{Co}}(E_{\mathrm{er}})$) are convolved to produce likelihood curves for the resulting $E_{\mathrm{thr}}$ of the four experiments considered here.  The results are shown in Table \ref{tab:e_thresholds}.  It is clear that all four experiments, even in the presence of the sharply falling $\mathcal{R}_e$ observed here, have sensitivity to all or part of the 2-5\,keV range favored by the DAMA results.

\begin{table}[hbp!]
\caption{\label{tab:e_thresholds} Four recent dark-matter searches using LXe: the second science run of ZEPLIN-III \cite{Akimov:2011tj}, results of XENON10 \cite{Angle:2007uj}, the recent 225\,live\,days reported from XENON100 \cite{Aprile:2012nq}, and the results of XMASS \cite{Abe:2012az}.  Shown are the applied electric fields used by each ($|\vec{\mathbf{E}}|$), their quoted scintillation thresholds ($S1_{\mathrm{thr}}$), their $^{57}$Co light yield ($LY_{\mathrm{Co}}$), and their electronic-recoil energy thresholds using this work ($E_{\mathrm{thr}}$).}
\begin{tabular}{l | r | c | c | c }
	\hline \hline
	\multirow{2}{*}{Experiment} & 
		\multirow{2}{*}{$|\vec{\mathbf{E}}|$\,(V/cm)} & 
		\multirow{2}{*}{$S1_{\mathrm{thr}}$ (PE)}& 
		\multirow{2}{*}{$LY_{\mathrm{Co}} (\frac{\mathrm{PE}}{\mathrm{keV}})$} & 
		\multirow{2}{*}{$E_{\mathrm{thr}}$\,(keV)} \\
		& & & & \\
	\hline
	\multirow{2}{*}{ZEPLIN-III} 	& 
		\multirow{2}{*}{3400~~~}	& 
		\multirow{2}{*}{2.6}		& 	
		\multirow{2}{*}{1.3}	& 
		\multirow{2}{*}{$2.8^{+0.5}_{-0.5}$} \\
		& & & & \\
	XENON10		& 
		730~~~		& 
		4.4		&	
		3.0	& 
		$2.5^{+0.4}_{-0.3}$ \\
	\multirow{2}{*}{XENON100}	& 
		\multirow{2}{*}{530~~~}		& 
		\multirow{2}{*}{3.0}		&	
		\multirow{2}{*}{2.3}	& 
		\multirow{2}{*}{$2.3^{+0.4}_{-0.3}$} \\
		& & & & \\		
	XMASS		& 
		0~~~		& 
		4.0	&  
		14.7 & 
		$1.1^{+0.4}_{-0.2}$ \\
	& & & & \\
	\hline \hline
\end{tabular}
\end{table}

\section{Summary}
\label{sec:summary}
The work presented here details a study of LXe's scintillation response to electronic recoils as low as 1.5\,keV.  The proportionality between deposited energy and scintillation signal, or ``light yield'' (LY), is observed to drop with decreasing energy beginning at $\sim$10\,keV, to a level roughly 40\% of its value at higher energies.  With the application of a static electric field of 450\,V/cm, we observe a reduction of the scintillation signal of roughly 75\% relative to the value at zero field, and see no significant energy dependence on this value between 1.5\,keV and 7.8\,keV.  With these values, we are able to extrapolate the electronic-recoil energy thresholds of the ZEPLIN-III \cite{Akimov:2011tj}, XENON10 \cite{Angle:2007uj}, XENON100 \cite{Aprile:2012nq}, and XMASS \cite{Abe:2012az} experiments.  These experiments report scintillation thresholds of 2.6\,PE, 4.4\,PE, 3.0\,PE, and 4.0\,PE, which, when applied with the results presented here, give energy thresholds of 2.8\,keV, 2.5\,keV, 2.3,keV, and 1.1\,keV, respectively.  We additionally investigate a discrepancy between the LY from the 9.4\,keV emission of $^{83\mathrm{m}}$Kr (which has in the past been considered for use as a standard calibration source) and other observed LY values nearby in energy.  We observe a time dependence of this scintillation signal which we show is most likely due to diffusion of electrons from the preceding 32.1\,keV decay of the same nucleus.  From this observation, we conclude that the 9.4\,keV peak is not an optimum standard calibration feature, and instead advocate the use of the 32.1\,keV for this purpose.

\begin{acknowledgments}
This work was supported by the Swiss National Foundation grant No.~200020-138225.  A.K.~and A.M.~received support from the University of Zurich Forschungskredit fellowship, Nos.~57161702 and 57161703, respectively.
\end{acknowledgments}

\end{document}